\documentstyle[12pt,epsf,epsfig]{article}
\newcount\timecount
\newcount\hours \newcount\minutes  \newcount\temp \newcount\pmhours

\hours = \time
\divide\hours by 60
\temp = \hours
\multiply\temp by 60
\minutes = \time
\advance\minutes by -\temp
\def\hour{\the\hours}
\def\minute{\ifnum\minutes<10 0\the\minutes
            \else\the\minutes\fi}
\def\clock{
\ifnum\hours=0 12:\minute\ AM
\else\ifnum\hours<12 \hour:\minute\ AM
      \else\ifnum\hours=12 12:\minute\ PM
            \else\ifnum\hours>12
                 \pmhours=\hours
                 \advance\pmhours by -12
                 \the\pmhours:\minute\ PM
                 \fi
            \fi
      \fi
\fi
}

\def\monthname{\relax\ifcase\month 0/\or January\or February\or
   March\or April\or May\or June\or July\or August\or September\or
   October\or November\or December\else\number\month/\fi}

\def\bold#1{\setbox0=\hbox{$#1$}%
     \kern-.025em\copy0\kern-\wd0
     \kern.05em\copy0\kern-\wd0
     \kern-.025em\raise.0433em\box0 }

\def\gappeq{\mathrel{\rlap {\raise.5ex\hbox{$>$}}
{\lower.5ex\hbox{$\sim$}}}}

\def\lappeq{\mathrel{\rlap{\raise.5ex\hbox{$<$}}
{\lower.5ex\hbox{$\sim$}}}}

\def\Yi{\eta^{i\ast}_{11} \left( \frac{y_{i}}{2} g' Z_{\chi_1} + 
        g T_{3i} Z_{\chi_2} \right) + \eta^{i\ast}_{12} 
        \frac{g m_{q_{i}} Z_{\chi_{5-i}}}{2 m_{W} B_{i}}}

\def\Xii{\eta^{i\ast}_{11} 
        \frac{g m_{q_{i}}Z_{\chi_{5-i}}^{\ast}}{2 m_{W} B_{i}} - 
        \eta_{12}^{i\ast} e_{i} g' Z_{\chi_1}^{\ast}}

\def\Wi{\eta_{21}^{i\ast}
        \frac{g m_{q_{i}}Z_{\chi_{5-i}}^{\ast}}{2 m_{W} B_{i}} -
        \eta_{22}^{i\ast} e_{i} g' Z_{\chi_1}^{\ast}}

\def\Vi{\eta_{22}^{i\ast} \frac{g m_{q_{i}} Z_{\chi_{5-i}}}{2 m_{W} B_{i}}
        + \eta_{21}^{i\ast}\left( \frac{y_{i}}{2} g' Z_{\chi_1}
        + g T_{3i} Z_{\chi_2} \right)}

\def\zthree{\delta_{1i} [g Z_{\chi_2} - g' Z_{\chi_1}]}
\def\zfour{\delta_{2i} [g Z_{\chi_2} - g' Z_{\chi_1}]}

\def\ga{\mathrel{\raise.3ex\hbox{$>$\kern-.75em\lower1ex\hbox{$\sim$}}}}
\def\la{\mathrel{\raise.3ex\hbox{$<$\kern-.75em\lower1ex\hbox{$\sim$}}}}
\def\gev{{\rm \, Ge\kern-0.125em V}}
\def\tev{{\rm \, Te\kern-0.125em V}}
\def\beq{\begin{equation}}
\def\eeq{\end{equation}}

\def\mchi{m_{\chi}}

\def\ohsq{\Omega_{\chi} h^2}

\def\m12{m_{1\!/2}}

\newcommand{\pb}{{\rm pb}}
\newcommand{\km}{{\rm km}}
\newcommand{\cm}{{\rm cm}}
\newcommand{\yr}{{\rm yr}}
\newcommand{\s}{{\rm s}}
\newcommand{\ethr}{E_{\rm th}}
\newcommand{\eopt}{E_{\rm opt}}

\newcommand{\postscript}[2]{\setlength{\epsfxsize}{#2\hsize}
   \centerline{\epsfbox{#1}}}

\begin{document}
\begin{titlepage}
\pagestyle{empty}
\baselineskip=21pt
\rightline{astro-ph/0110225}
\rightline{CERN--TH/2001-264}
\rightline{MIT--CTP--3181, UCI--TR--2001--26}
\rightline{UMN--TH--2026/01, TPI--MINN--01/44}
\vspace*{0.20in}
\begin{center}
{\large{\bf Prospects for Detecting Supersymmetric Dark Matter at 
Post-LEP Benchmark Points
}}
\end{center}
\begin{center}
\vskip 0.05in
{\bf John Ellis}$^1$,
{\bf Jonathan L.~Feng}$^{2,3}$,
{\bf Andrew Ferstl}$^4$,\\
{\bf Konstantin T.~Matchev}$^1$ and
{\bf Keith A.~Olive}$^{5}$
\vspace*{0.05in}

{\it
$^1${TH Division, CERN, CH--1211 Geneva 23, Switzerland}\\
$^2${Center for Theoretical Physics,\\
     Massachusetts Institute of Technology, Cambridge, MA 02139, USA}\\
$^3${Department of Physics and Astronomy, \\
     University of California, Irvine, CA 92697, USA}\\
$^4${Department of Physics, 
     Winona State University, Winona, MN 55987, USA}\\
$^5${Theoretical Physics Institute, School of Physics and Astronomy,\\
     University of Minnesota, Minneapolis, MN 55455, USA}\\
}
\vspace*{0.20in}
{\bf Abstract}
\end{center}
\baselineskip=18pt \noindent

A new set of supersymmetric benchmark scenarios has recently been
proposed in the context of the constrained MSSM (CMSSM) with universal
soft supersymmetry-breaking masses, taking into account the
constraints from LEP, $b \rightarrow s \gamma$ and $g_\mu - 2$. These
points have previously been used to discuss the physics reaches of
different accelerators. In this paper, we discuss the prospects for
discovering supersymmetric dark matter in these scenarios.  We
consider direct detection through spin-independent and spin-dependent
nuclear scattering, as well as indirect detection through relic
annihilations to neutrinos, photons, and positrons. We find that
several of the benchmark scenarios offer good prospects for direct
detection via spin-independent nuclear scattering and indirect
detection via muons produced by neutrinos from relic annihilations
inside the Sun, and some models offer good prospects for detecting
photons from relic annihilations in the galactic centre.

\vfill
\vskip 0.15in
\leftline{October 2001}
\end{titlepage}
\baselineskip=18pt

\section{Introduction}
\label{sec:introduction}

After the closure of LEP, at the start of Run II of the Tevatron
Collider, with the LHC experimental programme being prepared, and
linear $e^+ e^-$ collider projects under active discussion, now is an
appropriate time to review the available experimental constraints on
supersymmetry and assess the prospects for its discovery. In parallel
with present and future accelerator projects, many non-accelerator
experiments that may contribute to the search for supersymmetry are
underway or in preparation. These include direct searches for the
elastic scattering of astrophysical cold dark matter particles on
target nuclei, and indirect searches for particles produced by the
annihilations of supersymmetric relic particles inside the Sun or
Earth, in the galactic centre or in the galactic halo.

A set of benchmark supersymmetric model parameter choices was recently
proposed~\cite{Battaglia:2001zp} with the idea of exploring the
possible phenomenological signatures in different classes of
experiments in a systematic way. The proposed benchmark points were
chosen by first implementing the constraints on the CMSSM parameter
space \cite{EFGO} imposed by previous experiments, such as the
searches for sparticles \cite{SUSYWG} and Higgs bosons at LEP
\cite{LEPHiggs} and elsewhere, the measured rate for $b \to s \gamma$
decay \cite{bsgexpt}, and (optionally) the value of $g_\mu - 2$
recently reported by the BNL E821 experiment \cite{Brown:2001mg}. The
CMSSM parameter space was also constrained by requiring the calculated
supersymmetric relic density to fall within the range $0.1 <
\Omega_\chi h^2 < 0.3$ preferred by astrophysics and cosmology. Four
general regions of allowed parameter space were identified: a `bulk'
region at relatively low $m_0$ and $m_{1/2}$, a `focus point' region
\cite{Feng:2000mn,Feng:2000gh} at relatively large $m_0$, a
coannihilation `tail' extending out to relatively large $m_{1/2}$
\cite{EFOSi,glp}, and a possible `funnel' between the focus point and
coannihilation regions due to rapid annihilation via direct-channel
Higgs boson poles \cite{EFGOSi}.

The benchmark points were chosen not to provide an unbiased
statistical sample of the CMSSM parameter space, which is in any case
difficult to define in the absence of any unbiased {\it a priori}
measure, but rather to select representative examples of different
possibilities that cannot yet be logically excluded.  Note that while
these scenarios are confined to the context of supergravity, they span
a large range of dark matter properties.  While other
supersymmetry-breaking schemes lead to a variety of collider signals,
with respect to dark matter, they often predict vanishing or highly
suppressed thermal relic densities for the most natural candidate, the
neutralino.  These alternative scenarios therefore typically have no
viable dark matter candidates, at least without additional structure
and an accompanying loss of predictability.

\begin{figure}[tbp]
\begin{minipage}[t]{0.49\textwidth}
\postscript{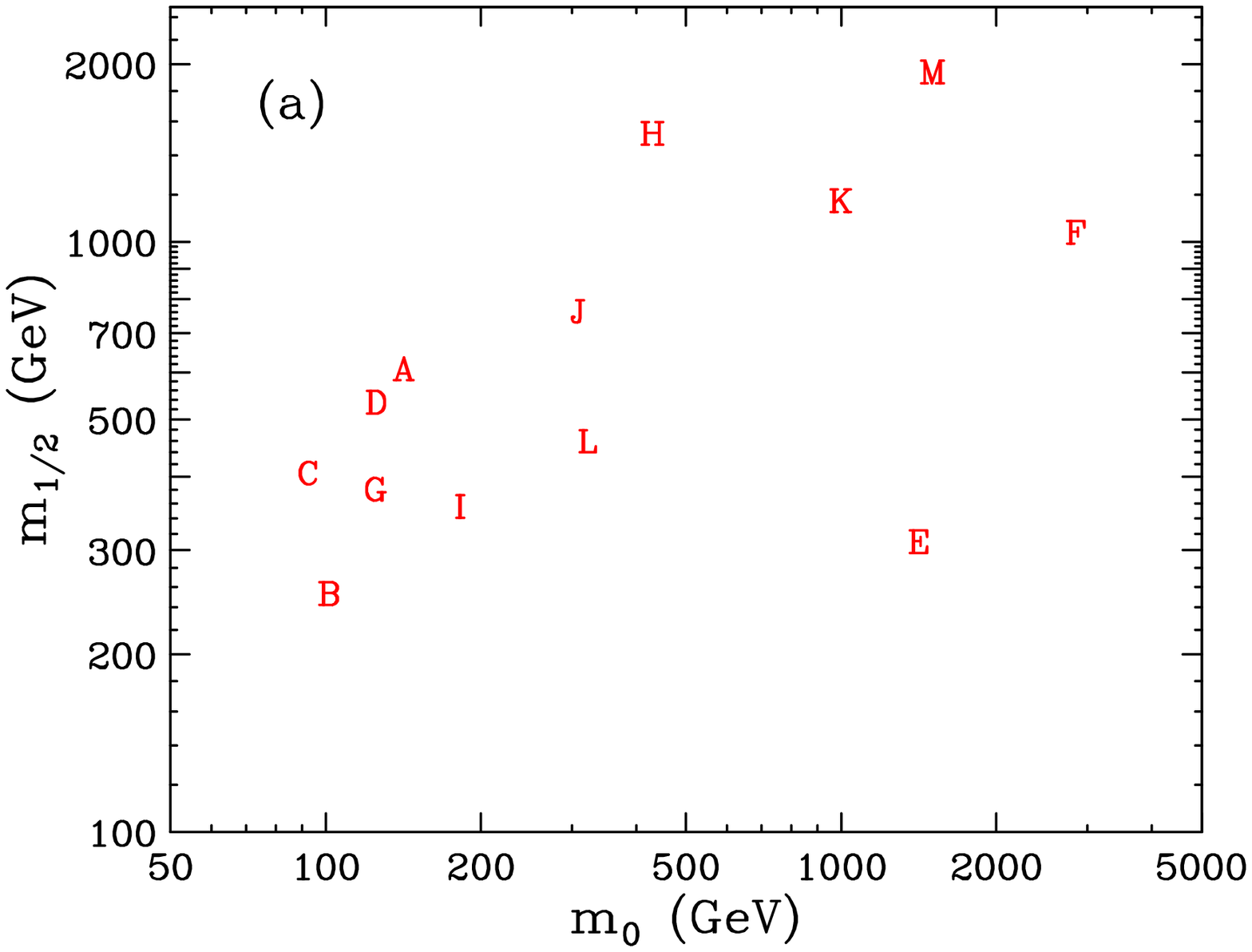}{0.99}
\end{minipage}
\hfill
\begin{minipage}[t]{0.49\textwidth}
\postscript{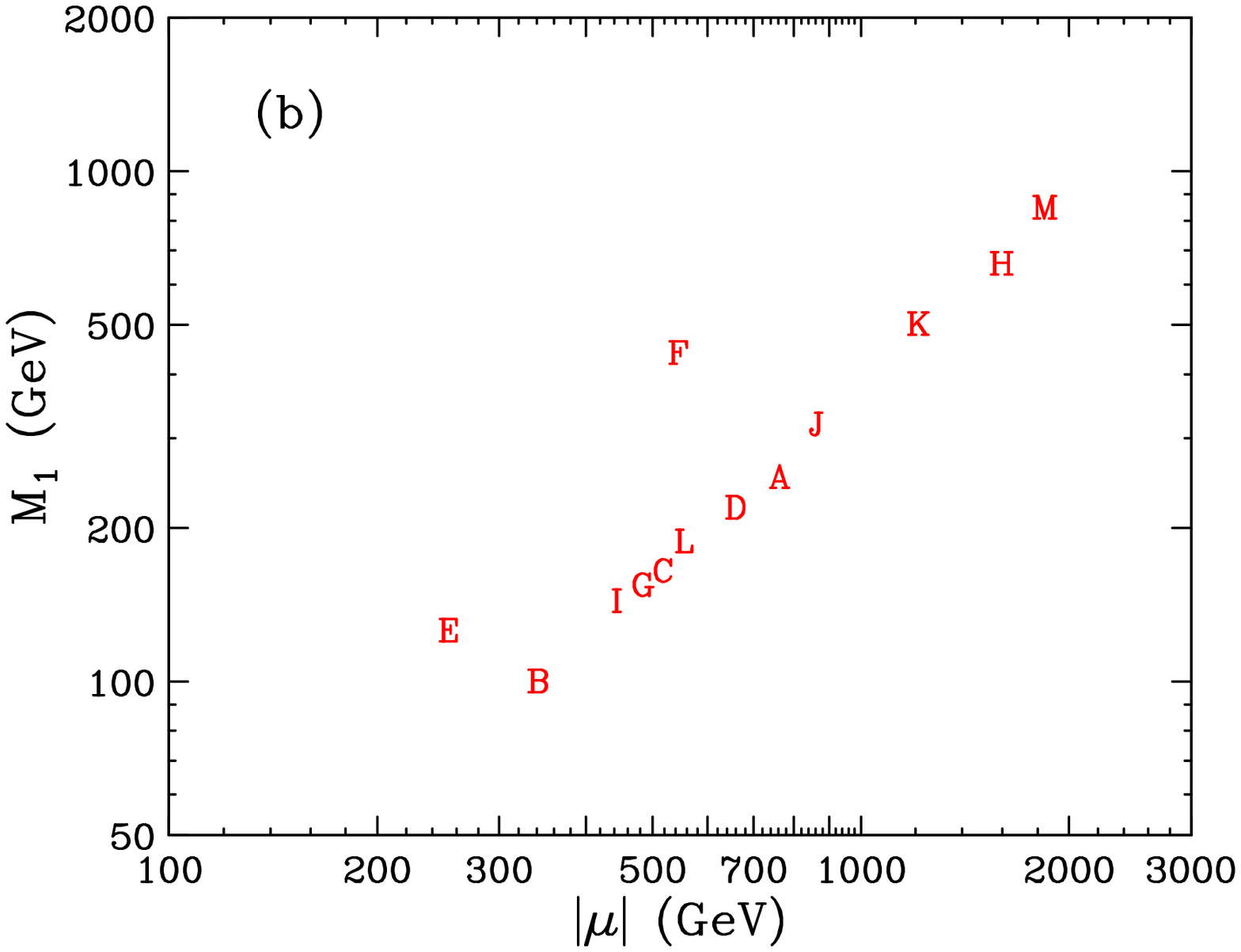}{0.99}
\end{minipage}
\caption{\label{fig:overview} {\it Benchmark 
points~\cite{Battaglia:2001zp} in the 
(a) $(m_0, m_{1/2})$ and (b) $(|\mu|, M_1)$ planes.  }}
\end{figure}

Of the 13 benchmark points, B, C, G, I, and L lie within the `bulk'
region; E and F are in the focus point region; A, D, H, and J are
strung out along the coannihilation tail; and K and M are chosen at
(relatively) large $m_{1/2}$ and $m_0$, in the rapid annihilation
funnel regions.  About half of the proposed points yield a value of
$g_\mu - 2$ within two standard deviations of the value reported by
BNL E821, but we did not impose this as an absolute requirement. For
example, two points with $\mu < 0$, the sign disfavoured by $g_\mu -
2$, were retained. Fig.~\ref{fig:overview} provides an overview of the
locations of the benchmark points in the $(m_0, m_{1/2})$ and $(|\mu|,
M_1)$ planes. We see that the proposed scenarios mainly have $m_{1/2}
> m_0$, except for the two focus point models E and F.  These also
have larger values of $M_1/|\mu|$, and therefore more Higgsino-like
lightest supersymmetric particles (LSPs).  Table~\ref{table:I}
displays many properties of the proposed scenarios, including the LSP
mass, its gaugino composition, its cosmological relic density, and
rates for the many astrophysical signatures to be discussed in
subsequent sections of this paper.

{\scriptsize
\begin{table}[p]
\centering
\begin{tabular}{|c||r|r|r|r|r|r|r|r|r|r|r|r|r|}
\hline
$\rule{0mm}{5mm}$
Model          & A   &  B  &  C  &  D  &  E  &  F  &  G  &  H  &  I  
&  J  &  K  &  L  &  M   \\ [2mm]
\hline
$\rule{0mm}{5mm}$
$m_{1/2}$      & 613 & 255 & 408 & 538 & 312 & 1043& 383 & 1537& 358 
& 767 & 1181& 462 & 1953 \\ [2mm]
$m_0$          & 143 & 102 &  93 & 126 & 1425& 2877& 125 & 430 & 188 
& 315 & 1000& 326 & 1500 \\ [2mm]
$\tan{\beta}$  & 5   & 10  &  10 & 10  & 10  & 10  & 20  & 20  & 35  
& 35  & 39.6& 45  & 45.6 \\ [2mm]
sgn($\mu$)     & $+$ & $+$ & $+$ & $-$ & $+$ & $+$ & $+$ & $+$ & $+$ 
& $+$ & $-$ & $+$ & $+$  \\ [2mm]
\hline
$\rule{0mm}{5mm}$
$m_{\chi}$
               &251.8& 98.1&163.8&221.0&119.2&434.2&153.7&663.6&143.1
&320.8&505.7&188.0&853.9 \\ [3mm]
$R_\chi$       &0.997&0.986&0.994&0.997&0.954&0.950&0.994&0.999&0.993
&0.998&0.999&0.995&0.999 \\ [3mm]
\hline
$\rule{0mm}{5mm}$
$\Omega h^2$   &0.26 &0.18 &0.14 &0.19 & 0.31&0.17 &0.16 &0.29 &0.16 
&0.20 &0.19 &0.21 &0.17  \\ [3mm]
\hline
$\rule{0mm}{5mm}$
$\sigma_P^{\rm sc}$ 
               &387.9&6567.&1031.&1.745&4859.&4121.&2262.&32.11&8953.
&335.3&0.061&5862.&32.61 \\ [3mm]
$\sigma_P^{\rm sp}$ 
               &0.260&11.06&1.622&0.518&102.4&14.15&2.236&0.022&3.045
&0.216&0.075&1.358&0.016 \\ [3mm]
$\sigma_N^{\rm sc}$ 
               &399.8&7002.&1085.&2.304&5004.&4221.&2426.&33.19&9730.
&357.5&0.192&6375.&34.04 \\ [3mm]
$\sigma_N^{\rm sp}$ 
               &0.224&8.750&1.331&0.434&64.19&8.831&1.805&0.017&2.416
&0.171&0.055&1.053&0.012 \\ [3mm]
\hline
$\rule{0mm}{5mm}$
$\Phi_\mu^\odot$
& 0.0138 & 5.43 & 0.706 & 0.0585 & 152. & 7.25 & 1.23 & $10^{-5}$  
& 1.809  & 0.0493 & 0.0089  & 1.002 & 0.0013 \\ [3mm]
$\Phi_\mu^\oplus$
&$10^{-9}$ &$10^{-5}$ &$10^{-7}$ &$10^{-13}$ &$10^{-5}$ &$10^{-5}$
&$10^{-6}$ &$10^{-12}$ &$10^{-4}$ &$10^{-8}$ &$10^{-13}$ &$10^{-4}$
 &$10^{-10}$ \\ [3mm]
\hline
$\rule{0mm}{5mm}$
$\Phi_\gamma^{1}$
&  1.428  & 84.29  & 10.19 & 2.248 & 85.59 & 39.60
&  63.90  & 0.204  & 535.0 & 25.86 & 119.4 & 992.4
&  37.48  \\ [3mm]
%
$\Phi_\gamma^{50}$
&  0.340  & 0.874  & 1.108  & 0.720  & 8.567  &  30.00  
&  5.065  & 0.450  & 31.25  & 17.37  & 180.7  &  160.0  
& 108.0 \\ [3mm]
\hline
$\rule{0mm}{5mm}$
$S/B$ 
& $10^{-7}$ 
& $10^{-5}$ 
& $10^{-6}$ 
& $10^{-7}$ 
& $10^{-3}$ 
& $10^{-4}$ 
& $10^{-6}$ 
& $10^{-9}$ 
& $10^{-6}$ 
& $10^{-8}$ 
& $10^{-9}$ 
& $10^{-8}$ 
& $10^{-10}$ 
\\ [3mm]
%
$E_{opt}$
&  153.6  & 50.04 & 83.56 & 130.4 & 60.79 & 264.8
&  78.37  & 338.4 & 73.00 & 202.1 & 298.4 & 95.89
&  315.9 \\ [3mm]
\hline
\end{tabular}
\caption{
\label{table:I} {\it Parameters and dark matter observables for
the benchmark points. The supersymmetric mass spectra are obtained
using {\tt ISASUGRA 7.51}~\cite{isasugra} with the listed input
parameters.  For all the benchmark points, we assume $A_0 = 0$ and
$m_t=175~\gev$.  All masses and energies are in GeV.  We define the
gaugino fraction of the lightest neutralino $\chi$ as $R_{\chi} \equiv
|Z_{\chi_1}|^2 + |Z_{\chi_2}|^2$, where $\chi = Z_{\chi_1} \tilde{B} +
Z_{\chi_2} \tilde{W}^0 + Z_{\chi_3} \tilde{H}^0_u + Z_{\chi_4}
\tilde{H}^0_d$.  The neutralino relic density $\ohsq$ is taken from
Table 2 of~\cite{Battaglia:2001zp}, and were calculated using {\tt
SSARD}~\cite{ssard}.  The spin-independent (spin-dependent) cross
sections on protons $\sigma^{\rm sc}_P$ ($\sigma^{\rm sp}_P$) and
neutrons $\sigma^{\rm sc}_N$ ($\sigma^{\rm sp}_N$) are calculated with
{\tt Neutdriver}~\cite{Jungman:1996df} and are given in units of
$10^{-12}~\pb$ ($10^{-6}~\pb$).  The muon fluxes from the Sun
($\Phi_\mu^\odot$) and the Earth ($\Phi_\mu^\oplus$) are in units of
$\km^{-2}~\yr^{-1}$.  The integrated photon fluxes $\Phi_\gamma^{1}$
($\Phi_\gamma^{50}$) for photon energy threshold $\ethr = 1~\gev$
($\ethr = 50~\gev$) are in units of $10^{-12}~\cm^{-2}~\s^{-1}$
($10^{-14}~\cm^{-2}~\s^{-1}$). Finally, $S/B$ is the maximal value of
the positron signal-to-background ratio, and $\eopt$ is the energy at
which this value is realized.}}
\end{table}
}

It was found previously \cite{Battaglia:2001zp} that, in the $g_\mu -
2$-friendly scenarios, supersymmetry was relatively easy to discover
and study at future colliders such as the LHC and a linear collider
with $E_{CM} = 1$~TeV, which would be able to observe rather
complementary subsets of CMSSM particles. However, some of the other
points might escape detection, except via observations of the lightest
neutral Higgs boson of the CMSSM. The most difficult points were
typically those in the focus point region, at the tip of the
coannihilation tail, or along the rapid-annihilation funnels, with
points F, H, and M being particularly elusive.

In this paper, we report on the prospects for the direct and indirect
detection of astrophysical dark matter for each of these benchmark
points.  We present cross sections for the elastic scattering of
supersymmetric relic particles off both protons and neutrons via both
spin-independent and spin-dependent matrix elements, the rates for
observing muons induced by the collisions in rock of energetic
neutrinos produced by relic annihilations inside the Sun and Earth,
the rates for photons produced by annihilations in the galactic
centre, and the rates for positrons produced by the annihilations of
relic particles in the galactic halo. In all cases, we take into
account the sensitivities of present and planned detectors in
estimating the observability of signals from relic particles. We
emphasize that all our results necessarily depend on the halo model
used: this is particularly true for the photon signal from the
galactic centre. This model-dependence enters when comparing the power
of various experimental probes.  However, for any given signature, our
conclusions concerning the relative ease with which different models
can be seen should be quite reliable.

The structure of this paper is as follows. In
Sec.~\ref{sec:constraints} we review briefly the experimental
constraints that were used as inputs when proposing the benchmark
points studied in this paper. In Sec.~\ref{sec:direct} we compare the
predictions of two different codes, {\tt
Neutdriver}~\cite{Jungman:1996df} and {\tt SSARD}~\cite{ssard}, for
direct dark matter detection, obtaining very similar results. We use
{\tt Neutdriver} to calculate muon rates from the Sun and Earth in
Sec.~\ref{sec:neutrinos}, and we follow the analysis
of~\cite{Feng:2001zu} to determine the photon and positron rates in
Secs.~\ref{sec:photons} and~\ref{sec:positrons},
respectively. Finally, in Sec.~\ref{sec:conclusions} we draw some
tentative conclusions about the detectability of dark matter particles
in the different allowed regions of parameter space, and we contrast
the prospects in accelerator and non-accelerator experiments.

\section{Constraints used to Select Benchmark Points}
\label{sec:constraints}

We restrict our attention to a constrained version of the MSSM (CMSSM)
which incorporates a minimal supergravity-inspired model of soft
supersymmetry breaking. Universal gaugino masses $m_{1/2}$, scalar
masses $m_0$ (including those of the Higgs multiplets) and trilinear
supersymmetry breaking parameters $A_0$ are used as inputs at the
supersymmetric grand unification scale. In this framework, the Higgs
mixing parameter $\mu$ can be derived (up to a sign) from the other
MSSM parameters by imposing the electroweak vacuum conditions for any
given value of $\tan \beta$.  Thus, given the set of input parameters
determined by $\{ m_{1/2}, m_0, A_0,\tan\beta, {\rm sgn}(\mu) \}$, the
entire spectrum of sparticles can be derived. For simplicity, we
further restrict our attention to $A_0 = 0$.

The available experimental and phenomenological constraints on the
CMSSM parameter space were implemented
in~\cite{Battaglia:2001zp}. These include the experimental constraints
obtained from searches for sparticles \cite{SUSYWG} and Higgs bosons
at LEP \cite{LEPHiggs}. In particular, attention was restricted to
parameter choices which guaranteed chargino masses $m_{\chi^\pm} >
103.5$ GeV~\cite{LEPSUSYWG_0103} and selectron masses $m_{\tilde e} >
99.4$ GeV~\cite{LEPSUSYWG_0101}. The lower limit on the mass of a
Standard Model Higgs boson imposed by the combined LEP experiments is
$113.5$~GeV~\cite{LEPHiggs}, and this limit also applies to the
lightest supersymmetric Higgs boson $h$ in the CMSSM.  To calculate
$m_h$ theoretically, we use the {\tt FeynHiggs}
code~\cite{Heinemeyer:2000yj}, which includes one-loop effects and
also the leading two-loop contributions. To account for uncertainties
in theoretical calculations of $m_h$~\cite{Heinemeyer:2000yj}, for any
given value of $m_t$, we restrict our CMSSM parameter choices to those
yielding $m_h \ge 113~{\rm GeV}$. In addition, the theoretical value
of $m_h$ in the MSSM is quite sensitive to $m_t$, the pole mass of the
top quark: we use $m_t = 175$~GeV as default. All but one of the
benchmark points satisfy $m_h > 113$~GeV.  In view of the expected
accuracy $\sim 3$~GeV of the {\tt FeynHiggs} code, we consider that
all the proposed points are compatible with the LEP lower limit of
$113.5$~GeV~\cite{LEPHiggs}.

We also compute the rate for $b \to s \gamma$ decay and compare it
with the experimental range \cite{bsgexpt}.  We implement the new NLO
$b \to s \gamma$ calculations of~\cite{newbsgcalx} when ${\tilde M} >
500$~GeV, where ${\tilde M} = {\rm min}(m_{\tilde q}, m_{\tilde
g})$. Otherwise, we use only the LO calculations and assign a larger
theoretical error. For the experimental value, we combine the CLEO
measurement with the recent BELLE result~\cite{bsgexpt}, ${\cal B} (b
\to s \gamma) = (3.21 \pm 0.44 \pm 0.26) \times 10^{-4}$. In our
implementation, we allow CMSSM parameter choices that, after including
the theoretical errors $\sigma_{th}$ due to the scale and model
dependences, may fall within the 95\% confidence level range $2.33
\times 10^{-4} < {\cal B}(b \to s \gamma) < 4.15 \times 10^{-4}$.

The final experimental contraint we consider is the $g_\mu - 2$ value
reported by the BNL E821 experiment~\cite{Brown:2001mg}. This
experiment has found an apparent discrepancy with the Standard Model
prediction at the level of 2.6 $\sigma$: $\delta a_\mu \; = \; (43 \pm
16) \times 10^{-10}.$ A large number of theoretical papers have
discussed the interpretation of the BNL measurement within
supersymmetry~\cite{Everett:2001tq,ENO}, and they generally agree that
$\mu > 0$ is favoured by the BNL measurement. The calculations we use
in this paper are taken from~\cite{ENO}, which are based on~\cite{IN},
including also the leading two-loop electroweak correction
factor~\cite{CM2l}.

We assume that $R$ parity is conserved, and that the stable LSP is the
lightest neutralino $\chi$ \cite{EHNOS}. We then constrain the CMSSM
parameter space by requiring the calculated supersymmetric relic
density to fall within the range $0.1 < \Omega_\chi h^2 < 0.3$
preferred by astrophysics and cosmology.  The upper limit on
$\Omega_\chi h^2$ is conservative, being based only on the lower limit
on the age of the Universe of 12 Gyr.  Smaller values of $\Omega_\chi
h^2$ are certainly possible, since some of the cold dark matter might
not consist of LSPs.  However, allowing smaller values of $\Omega_\chi
h^2$ would open up only a very small extra region of the $(m_0,
m_{1/2})$ plane.

Good overall consistency was found in~\cite{Battaglia:2001zp} between
these relic density calculations, the LEP and other sparticle mass
limits, the LEP Higgs limit, measurements of $b \to s \gamma$ and the
recent BNL measurement of $g_\mu - 2$, if $\mu > 0$ and $\tan \beta
\gappeq 5$.  For $\tan\beta \ga 50$, there are not substantial regions
with consistent electroweak vacua.

The values of the CMSSM parameters for the benchmark points are shown
in Table~\ref{table:I}.  {}From these, soft masses are determined with
{\tt ISASUGRA 7.51}, and relic densities are calculated with a recent
analysis~\cite{EFGOSi} using {\tt SSARD} that extends previous
results~\cite{EFGO} to larger $\tan \beta > 20$.  The chosen values of
$\tan \beta$ range from 5 to about $50$. In deference to $g_\mu - 2$,
most of the points proposed have $\mu > 0$, but only about a half of
the chosen points yield values of $g_\mu - 2$ within $2\sigma$ of the
present central experimental value, and two of the points have $\mu <
0$.  The amount of CMSSM parameter fine-tuning required for
electroweak symmetry breaking, along with the sensitivity of the relic
density to the precise values of the input CMSSM parameters, are given
in~\cite{Battaglia:2001zp} together with the corresponding sparticle
spectra.

\section{Direct Detection via Elastic Scattering}
\label{sec:direct}

The prospects for direct detection of neutralinos can be reduced to
the computation of the neutralino-proton elastic scattering cross
section.  We first review the ingredients of this calculation that are
implemented in {\tt SSARD}.  The MSSM Lagrangian leads to the
following low-energy effective four-fermion Lagrangian suitable for
describing elastic $\chi$-nucleon scattering~\cite{FFO1}:
\begin{eqnarray}
{\cal L} & = & \bar{\chi} \gamma^\mu \gamma^5 \chi \bar{q_{i}} 
\gamma_{\mu} (\alpha_{1i} + \alpha_{2i} \gamma^{5}) q_{i} +
\alpha_{3i} \bar{\chi} \chi \bar{q_{i}} q_{i} \nonumber \\
& + & \alpha_{4i} \bar{\chi} 
\gamma^{5} \chi \bar{q_{i}} \gamma^{5} q_{i}+
\alpha_{5i} \bar{\chi} \chi \bar{q_{i}} \gamma^{5} q_{i} +
\alpha_{6i} \bar{\chi} \gamma^{5} \chi \bar{q_{i}} q_{i} \ .
\label{lagr}
\end{eqnarray}
This Lagrangian is to be summed over the quark generations, and the
subscript $i$ labels up-type quarks ($i=1$) and down-type quarks
($i=2$).  The terms with coefficients $\alpha_{1i}, \alpha_{4i},
\alpha_{5i}$ and $\alpha_{6i}$ are velocity-dependent contributions
and may be neglected for the purpose of direct detection calculations.
The coefficients relevant for our discussion are, then, the
spin-independent or scalar coefficients
\begin{eqnarray}
\alpha_{3i} & = & - \frac{1}{2(m^{2}_{1i} - m^{2}_{\chi})} Re \left[
\left( X_{i} \right) \left( Y_{i} \right)^{\ast} \right] 
- \frac{1}{2(m^{2}_{2i} - m^{2}_{\chi})} Re \left[ 
\left( W_{i} \right) \left( V_{i} \right)^{\ast} \right] \nonumber \\
& & \mbox{} - \frac{g m_{qi}}{4 m_{W} B_{i}} \left[ Re \left( 
\zthree \right) D_{i} C_{i} \left( - \frac{1}{m^{2}_{H_{1}}} + 
\frac{1}{m^{2}_{H_{2}}} \right) \right. \nonumber \\
& & \mbox{} +  Re \left. \left( \zfour \right) \left( 
\frac{D_{i}^{2}}{m^{2}_{H_{2}}}+ \frac{C_{i}^{2}}{m^{2}_{H_{1}}} 
\right) \right] \ ,
\label{alpha3}
\end{eqnarray}
and the  spin-dependent coefficients
\begin{eqnarray}
\alpha_{2i} & = & \frac{1}{4(m^{2}_{1i} - m^{2}_{\chi})} \left[
\left| Y_{i} \right|^{2} + \left| X_{i} \right|^{2} \right] 
+ \frac{1}{4(m^{2}_{2i} - m^{2}_{\chi})} \left[ 
\left| V_{i} \right|^{2} + \left| W_{i} \right|^{2} \right] \nonumber \\
& & \mbox{} - \frac{g^{2}}{4 m_{Z}^{2} \cos^{2}{\theta_{W}}} \left[
\left| Z_{\chi_3} \right|^{2} - \left| Z_{\chi_4} \right|^{2}
\right] \frac{T_{3i}}{2} \ .
\label{alpha2}
\end{eqnarray}
Here $m_{1i}$ and $m_{2i}$ are the squark mass eigenvalues,   
\begin{eqnarray}
X_{i}& \equiv& \Xii \nonumber \\
Y_{i}& \equiv& \Yi \nonumber \\
W_{i}& \equiv& \Wi \nonumber \\
V_{i}& \equiv& \Vi \ ,
\label{xywz}
\end{eqnarray}
and the coefficients $Z_{\chi_i}$ define the composition of the
lightest neutralino through
\begin{equation}
\chi = Z_{\chi_1}\tilde{B} + Z_{\chi_2}\tilde{W} +
Z_{\chi_3}\tilde{H_{1}} + Z_{\chi_4}\tilde{H_{2}} \ .
\label{id}
\end{equation}
The parameters $e_i, T_{3i}, y_i$ denote electric charge, isospin and
hypercharge (normalized so that $e_i = T_{3i} + {y_i\over 2}$),
respectively, and
\begin{eqnarray}
\delta_{1i} = (Z_{\chi_3}, Z_{\chi_4}) && \delta_{2i} = 
(Z_{\chi_4},-Z_{\chi_3})
\nonumber \\
B_{i} = (\sin{\beta}, \cos{\beta}) && A_{i} = (\cos{\beta} , 
-\sin{\beta}) 
\nonumber \\
C_{i} = (\sin{\alpha} , \cos{\alpha}) && D_{i} = (\cos{\alpha} , 
- \sin{\alpha}) 
\label{moredefs}
\end{eqnarray}
for (up, down) type quarks. We denote by $m_{H_2} < m_{H_1}$ the two
scalar Higgs masses, and $\alpha$ is the Higgs mixing angle. Finally,
$\eta^i_{jk}$ are elements of the matrix that diagonalizes squark mass
matrices through ${\rm diag}(m^2_{1i}, m^2_{2i}) \equiv \eta^i M_i^2
(\eta^i)^{-1}$.

The spin-independent (scalar) part of the cross section can be written
as
\begin{equation}
\sigma_{3} = \frac{4 m_{r}^{2}}{\pi} \left[ Z f_{p} + (A-Z) f_{n} 
\right]^{2} \ ,
\label{si}
\end{equation}
where $m_{r}$ is the reduced neutralino mass and $A,Z$ are the atomic
number and nuclear electric charge,
\begin{equation}
\frac{f_{p}}{m_{p}} = \sum_{q=u,d,s} f_{Tq}^{(p)} 
\frac{\alpha_{3q}}{m_{q}} +
\frac{2}{27} f_{TG}^{(p)} \sum_{c,b,t} \frac{\alpha_{3q}}{m_q} \ ,
\label{f}
\end{equation}
where $m_p$ is the proton mass, and $f_n$ is defined similarly.  The
parameters $f_{Tq}^{(p)}$ are defined by
\begin{equation}
m_p f_{Tq}^{(p)} \equiv \langle p | m_{q} \bar{q} q | p \rangle
\equiv m_q B_q \ ,
\label{defbq}
\end{equation}
while $ f_{TG}^{(p)} = 1 - \sum_{q=u,d,s} f_{Tq}^{(p)} $~\cite{SVZ}.
Following the analysis in \cite{EFlO1,EFlO2} we  
use the following values of 
$f_{Tq}^{(p)}$:
\begin{eqnarray}
f_{Tu}^{(p)} = 0.020 \pm 0.004\ , && f_{Td}^{(p)} = 0.026 \pm 0.005
\nonumber \\
\qquad f_{Ts}^{(p)} = 0.118 \pm 0.062 \ , &&
\label{pinput}
\end{eqnarray}
where essentially all the error in $f_{Ts}^{(p)}$ arises from the
uncertainty in strangeness composition 
\beq 
y \equiv {2 B_s \over B_d
+ B_u} = 0.2 \pm 0.1 \ .
\label{defy}
\eeq 
The corresponding values for the neutron are
\begin{eqnarray}
f_{Tu}^{(n)} = 0.014 \pm 0.003\ , && f_{Td}^{(n)} = 0.036 \pm 0.008
\nonumber \\
\qquad f_{Ts}^{(n)} = 0.118 \pm 0.062 \ . &&
\label{ninput}
\end{eqnarray}
These values are based in part on the experimental value of the
$\pi$-nucleon $\sigma$ term~\cite{Gasser}
\beq
\sigma \equiv {1 \over 2} (m_u + m_d) \times (B_d + B_u) = 45 \pm
8~{\rm MeV} \ .
\label{sigma}
\eeq 
The larger value of $\sigma = 65$ MeV \cite{ol} considered by
\cite{arn2} leads to scattering cross section which are larger by a
factor of about 3.  It is clear already that the difference between
the scalar parts of the cross sections for scattering off protons and
neutrons must be rather small.

The spin-dependent part of the elastic $\chi$-nucleus cross section
can be written as
\begin{equation}
\sigma_{2} = \frac{32}{\pi} G_{F}^{2} m_{r}^{2} \Lambda^{2} J(J + 1) \
, 
\label{sd}
\end{equation}
where $G_F$ is the Fermi constant, $m_r$ is again the reduced
neutralino mass, $J$ is the spin of the nucleus, and
\begin{equation}
\Lambda \equiv \frac{1}{J} (a_{p} \langle S_{p} \rangle + a_{n} 
\langle S_{n} \rangle) \ ,
\label{lamda}
\end{equation}
where $\langle S_{p,n} \rangle $ are expectation values of the spin
content of the nucleus and
\begin{equation}
a_{p} = \sum_{i} \frac{\alpha_{2i}}{\sqrt{2} G_{F}} \Delta_{i}^{(p)} \
, \quad 
a_{n} = \sum_{i} \frac{\alpha_{2i}}{\sqrt{2} G_{F}} \Delta_{i}^{(n)} 
\ .
\label{a}
\end{equation}
The factors $\Delta_{i}^{(p,n)}$ parameterize the quark spin content
of the nucleon. A recent global analysis of QCD sum rules for the
$g_1$ structure functions~\cite{Mallot:2000qb}, including ${\cal
O}(\alpha_s^3)$ corrections, corresponds formally to the values
\begin{eqnarray}
\Delta_{u}^{(p)} = 0.78 \pm 0.02\ , & \Delta_{d}^{(p)} = -0.48 \pm
0.02 \nonumber \\
 \Delta_{s}^{(p)} = - 0.15 \pm 0.02\ . &
\label{spincontent}
\end{eqnarray}
In the case of the neutron, we have
$\Delta_{u}^{(n)} = \Delta_{d}^{(p)}, 
\Delta_{d}^{(n)} = \Delta_{u}^{(p)}$, and 
$\Delta_{s}^{(n)} = \Delta_{s}^{(p)}$.

The calculation of the neutralino-nucleon elastic scattering
cross-sections in {\tt Neutdriver} is based on~\cite{Drees:1993bu}.
The calculation of the spin-dependent contribution is identical to the
one presented above. However, the spin-independent computation
contains several additional pieces. First, the heavy flavor squark
contribution is treated in exact one-loop approximation as in
\cite{Drees:1993bu}, and (\ref{f}) is replaced by
\begin{equation}
\frac{f_{p}}{m_{p}} = \sum_{q=u,d,s} f_{Tq}^{(p)} 
\frac{\alpha_{3q}^{\tilde q}}{m_{q}} 
-\frac{8\pi}{9\alpha_S} f_{TG}^{(p)} 
\left[ B_D - \frac{\mchi^2}{4} B_{1D} \right],
\label{fnew}
\end{equation}
where $B_D$ and $B_{1D}$ are loop integrals defined in Eqs.~(18a) and
(18c) of~\cite{Drees:1993bu}, respectively.  Second, {\tt Neutdriver}
also includes a Higgs contribution through squark loops (see the last
term in Eq.~(43) of \cite{Drees:1993bu}).  Finally, it includes
several contributions from twist-2 operators, which are listed in
Eq.~(46) of \cite{Drees:1993bu}.

\begin{figure}[t]
\begin{minipage}[t]{0.49\textwidth}
\postscript{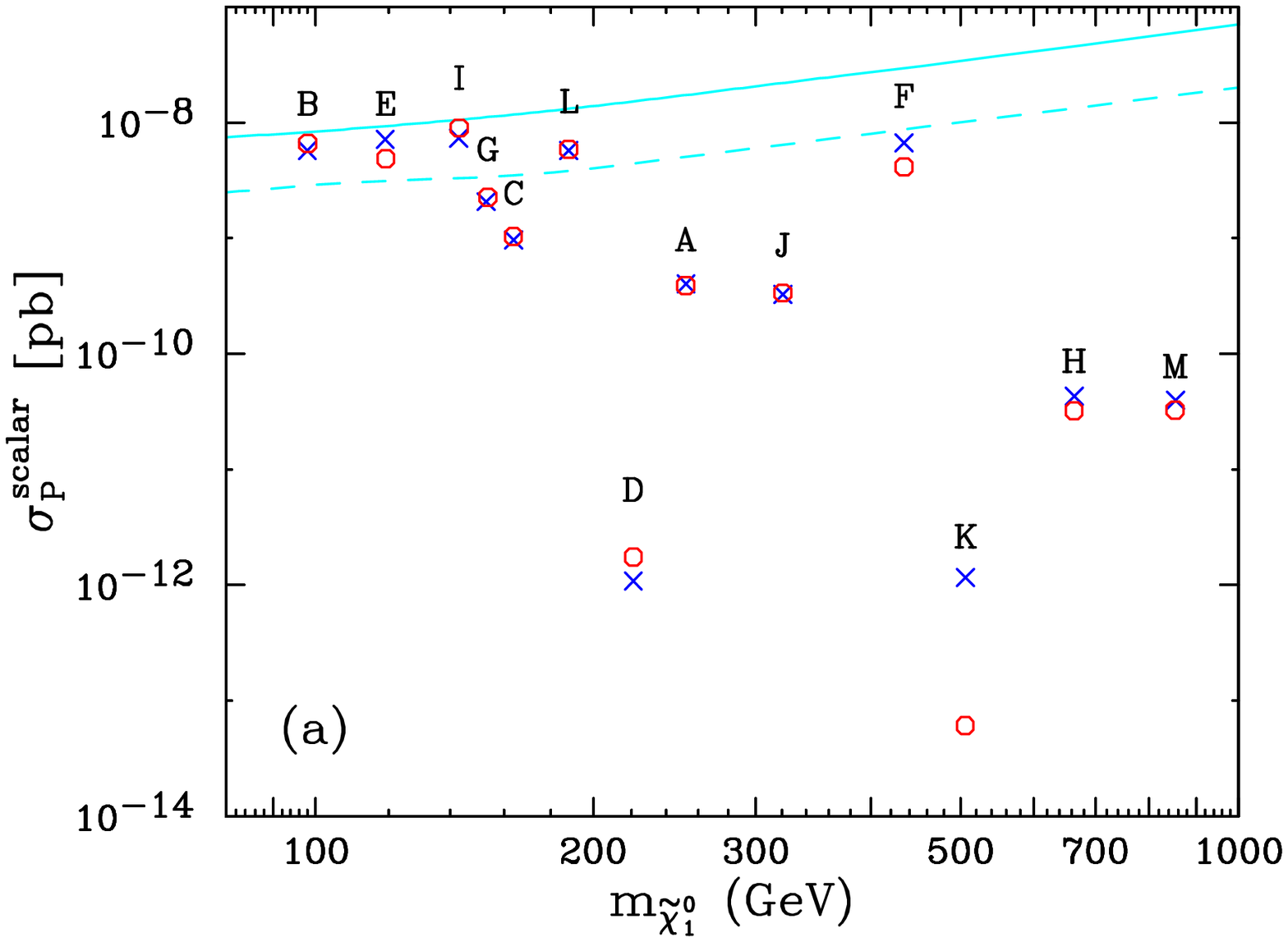}{0.99}
\end{minipage}
\hfill
\begin{minipage}[t]{0.49\textwidth}
\postscript{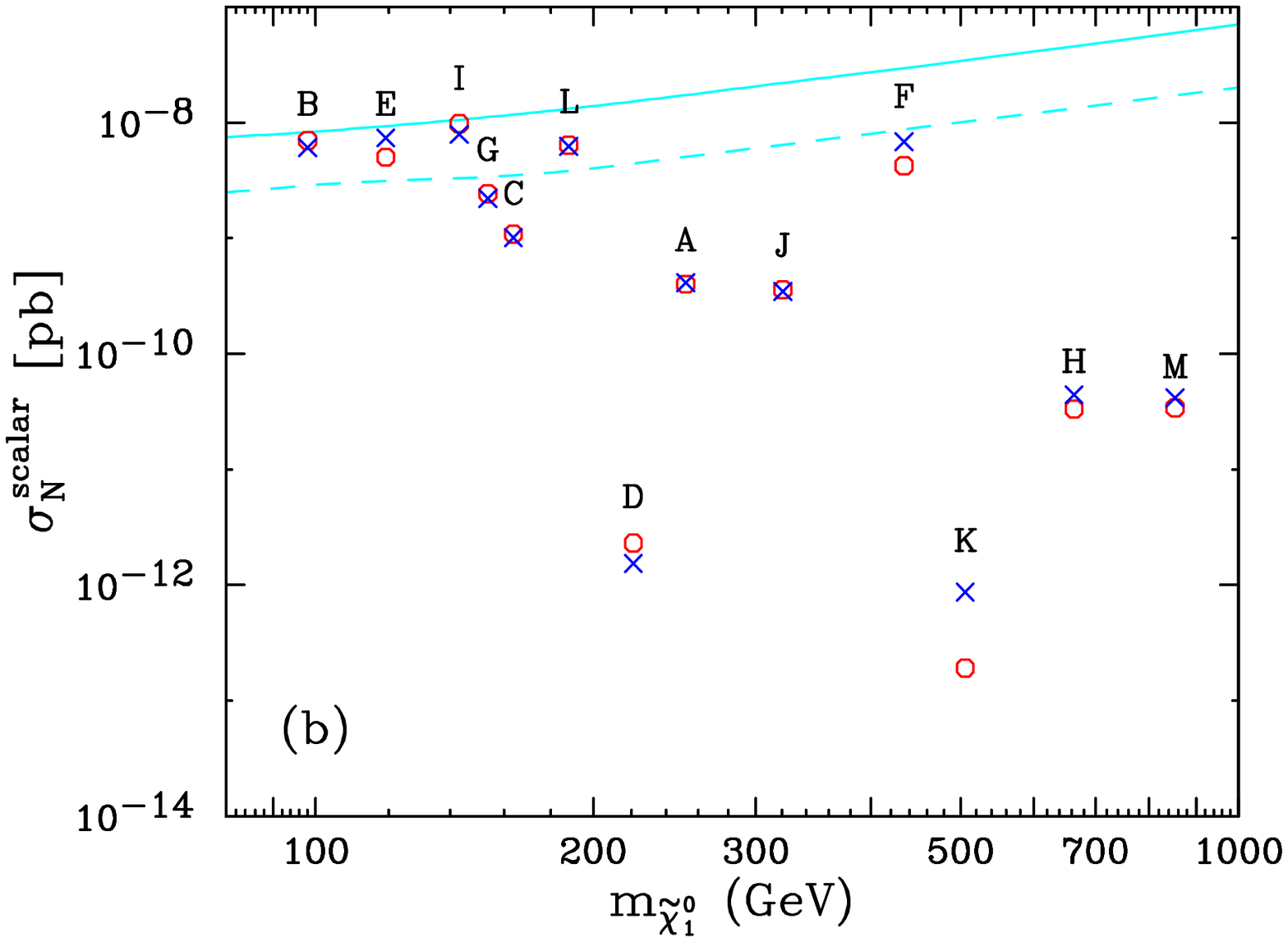}{0.99}
\end{minipage}
\\
\begin{minipage}[t]{0.49\textwidth}
\postscript{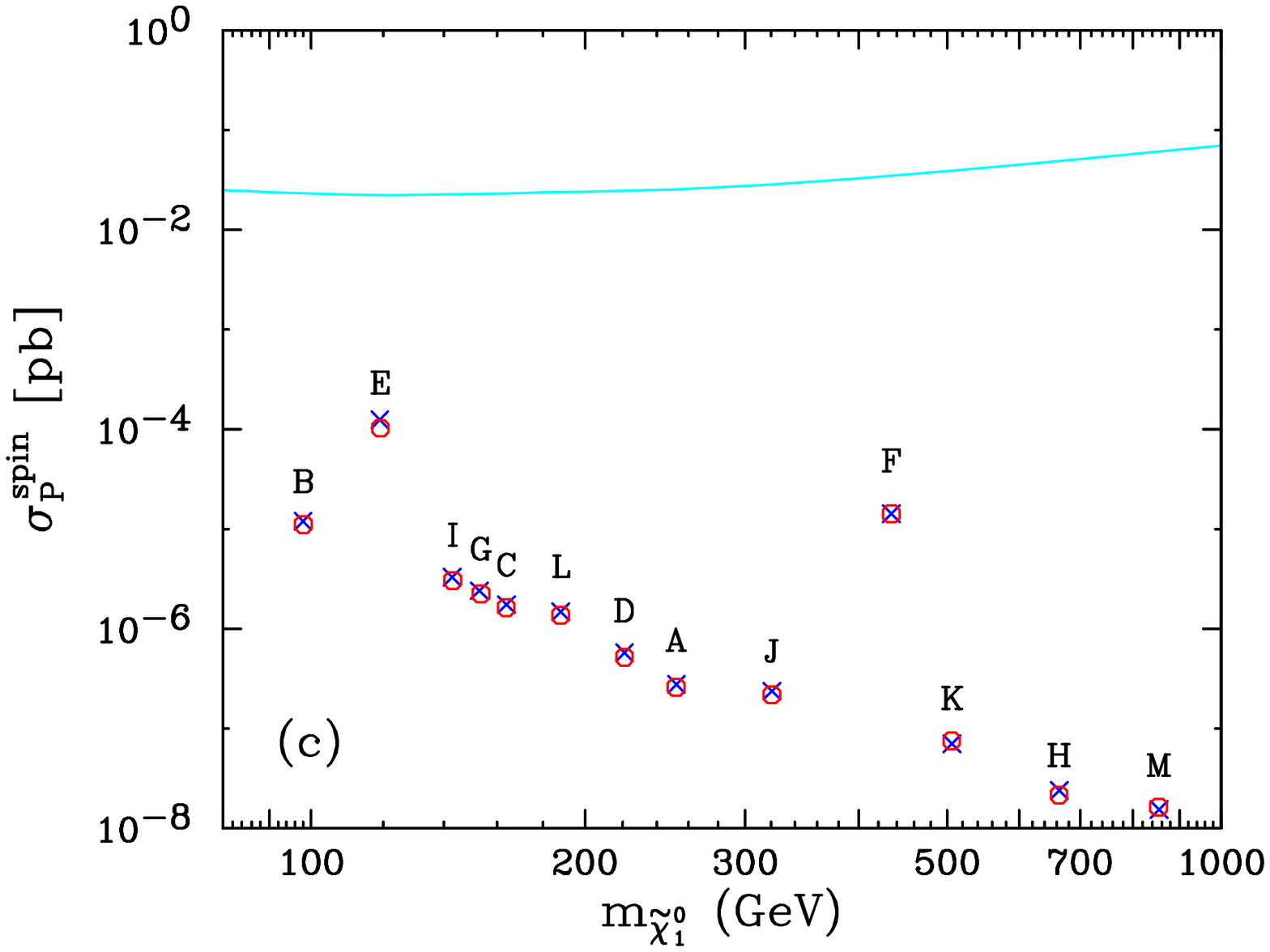}{0.99}
\end{minipage}
\hfill
\begin{minipage}[t]{0.49\textwidth}
\postscript{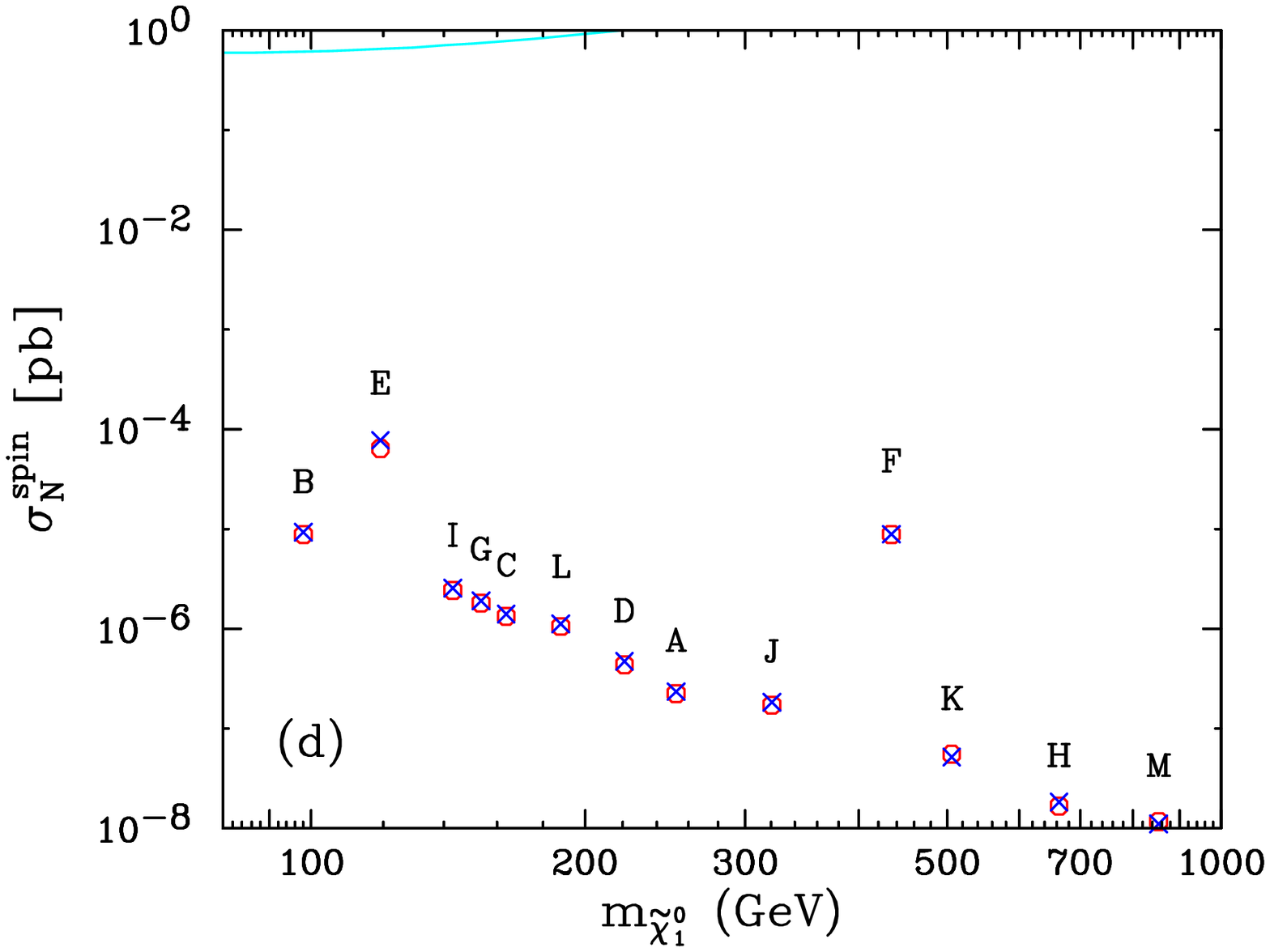}{0.99}
\end{minipage}
\caption{\label{fig:direct} {\it Elastic cross sections for (a,b)
spin-independent scattering and (c,d) spin-dependent scattering on
(a,c) protons and (b,d) neutrons. The predictions of {\tt SSARD} (blue
crosses) and {\tt Neutdriver} (red circles) for neutralino-nucleon
scattering are compared.  Projected sensitivities (a,b) for CDMS
II~\cite{Schnee:1998gf} and CRESST~\cite{Bravin:1999fc} (solid) and
GENIUS~\cite{GENIUS} (dashed) and (c) for a 100 kg NAIAD
array~\cite{Spooner:2000kt}, as well as (d) the existing DAMA
limit~\cite{Bernabei:1998ad} are also shown.}}
\end{figure}

Numerical values from {\tt Neutdriver} for the spin-independent and
the spin-dependent components of the elastic cross sections for the
scattering of neutralinos on protons and neutrons for each of the
benchmark points are presented in Table~\ref{table:I}.  (For other
recent work in the CMSSM, see, {\em e.g.}, \cite{Lahanas:2001mu}.)  In
Fig.~\ref{fig:direct}, we compare the results for the spin-independent
$\sigma_P^{sc}$ and spin-dependent $\sigma_P^{sp}$ cross-sections for
neutralino-proton and neutralino-neutron scattering using {\tt
SSARD}~\cite{ssard} and {\tt Neutdriver}~\cite{Jungman:1996df}. (For
the latter, we have changed the default values of the quantities
$f_{Tq}^{(p)}$, $f_{Tq}^{(n)}$, $\Delta_q^{(p)}$, and $\Delta_q^{(n)}$
to match those in (\ref{pinput}), (\ref{ninput}) and
(\ref{spincontent}).) The differences are insignificant relative to
the effects of different choices of CMSSM model parameters.  Recall
also that the mass spectra outputs of {\tt SSARD} and {\tt ISASUGRA}
differ, as may be seen by comparing Tables 1 and 3
of~\cite{Battaglia:2001zp}.  Fig.~\ref{fig:direct} shows the projected
sensitivities (a,b) for CDMS II~\cite{Schnee:1998gf} and
CRESST~\cite{Bravin:1999fc} (solid) and GENIUS~\cite{GENIUS} (dashed),
and (c) a 100 kg NAIAD~\cite{Spooner:2000kt} detector, as well as (d)
the existing DAMA limit~\cite{Bernabei:1998ad}. Obtaining a
competitive limit for the spin-dependent scattering on a neutron in
the latter case might be possible with a large $^{73}$Ge or Xenon
detector.

As was found in \cite{EFlO1}, there are strong cancellations in the
spin-independent cross sections when $\mu < 0$.  These cancellations
are due to sign differences between the up- and down-type quark
contributions to the Higgs exchange terms in $\alpha_3$ in
(\ref{alpha3}).  Nominally, these cancellations occur only for a
specific range in the neutralino mass.  For $\tan \beta = 10$, the
cancellations occur for $m_\chi\simeq 150 - 350$ GeV, and are
particularly effective when $m_\chi \simeq 200 - 250$ GeV. As one can
see in Table~\ref{table:I} and Fig.~\ref{fig:direct}, point D falls
exactly into this range, thus explaining why its scalar cross section
is anomalously small.  Similarly, for $\tan \beta = 35$, there are
strong cancellations at $m_\chi \simeq 400 - 600$ GeV
\cite{EFlO3}. Unfortunately, point K happens to fall in this range as
well.  Thus the two benchmark points with $\mu < 0 $ are predicted to
have very small spin-independent cross sections, but this would not
generally be true for other CMSSM models with $\mu < 0$. As one might
expect, the differences between the {\tt SSARD} and {\tt Neutdriver}
codes are largest for these points that exhibit delicate
cancellations.

Comparing the benchmark model predictions with the projected
sensitivities, we see that spin-independent scattering seems to offer
the best prospects for direct detection.  Among the proposed benchmark
points, models I, B, E, L, G, F, and C seem to offer the best
detection prospects. In particular, the first four of these models
would apparently be detectable with the proposed GENIUS detector.

\section{Neutrinos from Annihilations in the Sun and Earth}
\label{sec:neutrinos}

Dark matter particles collect in the gravitational wells at the
centers of astrophysical bodies, leading to large densities and
enhanced pair annihilation rates. While most annihilation products are
immediately trapped or absorbed, neutrinos may propagate for long
distances and be detected near the Earth's surface through their
charged-current conversion to muons.  High-energy muons produced by
neutrinos from the centers of the Sun~\cite{Silk:1985ax} and
Earth~\cite{Freese:1986qw} are therefore prominent signals for
indirect dark matter detection.

The muon detection rate is dependent on both the neutralino
annihilation rate and the resulting neutrino energy spectrum.  The
neutralino annihilation rate is proportional to the present dark
matter density at the core of the Sun or Earth.  Determinations of
these densities are involved, but well understood.  Various aspects of
these calculations are reviewed in~\cite{Jungman:1996df}, and
estimates of neutralino annihilation rates in the CMSSM for both the
Sun and the Earth are given in~\cite{Feng:2001zu}. (For other recent
work in the CMSSM, see, {\em e.g.}, \cite{neutrinos,Barger:2001ur}.)
For the Sun, the annihilation rate has typically reached equilibrium
and decreases for increasing neutralino mass.

The neutrino energy spectrum depends on the neutralino composition.
Neutralinos annihilate primarily to fermion pairs and gauge boson
pairs.  Annihilation to fermion pairs is helicity-suppressed, and so
is significant only for heavy fermions, such as $b$ quarks and $\tau$
leptons, and $t$ quarks if kinematically allowed.  Neutrinos from
these decays are typically rather soft.  Annihilation to gauge bosons
is possible only for neutralinos that are heavier than $W$ bosons and
have a significant Higgsino component.  When possible, however, these
annihilation channels typically dominate, producing hard neutrinos
from two-body gauge boson decay.  In this case, the muon flux is
greatly enhanced, as both the cross section for conversion to muons
and the muon range are proportional to the neutrino energy.

\begin{figure}[tb]
\begin{minipage}[t]{0.49\textwidth}
\postscript{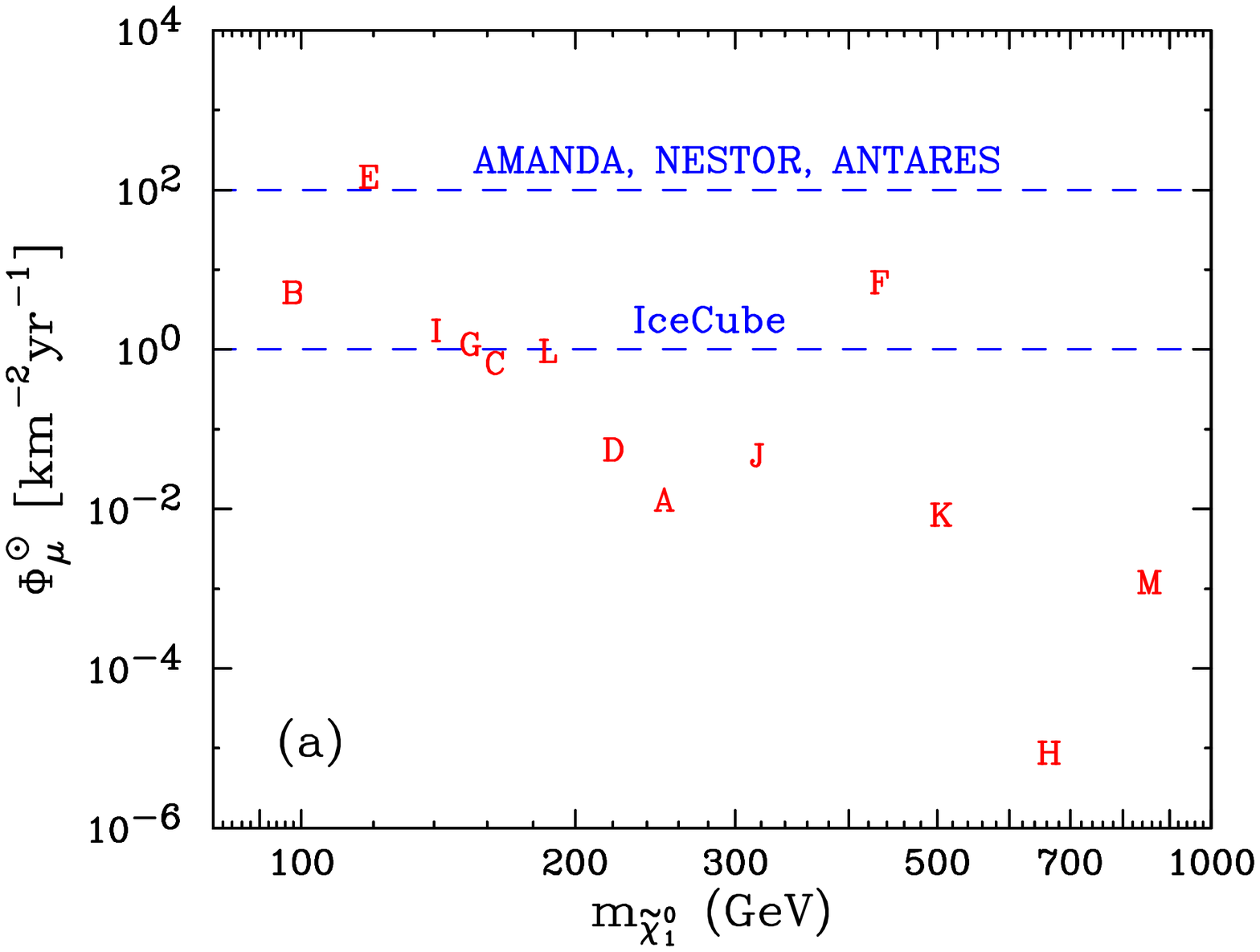}{0.99}
\end{minipage}
\hfill
\begin{minipage}[t]{0.49\textwidth}
\postscript{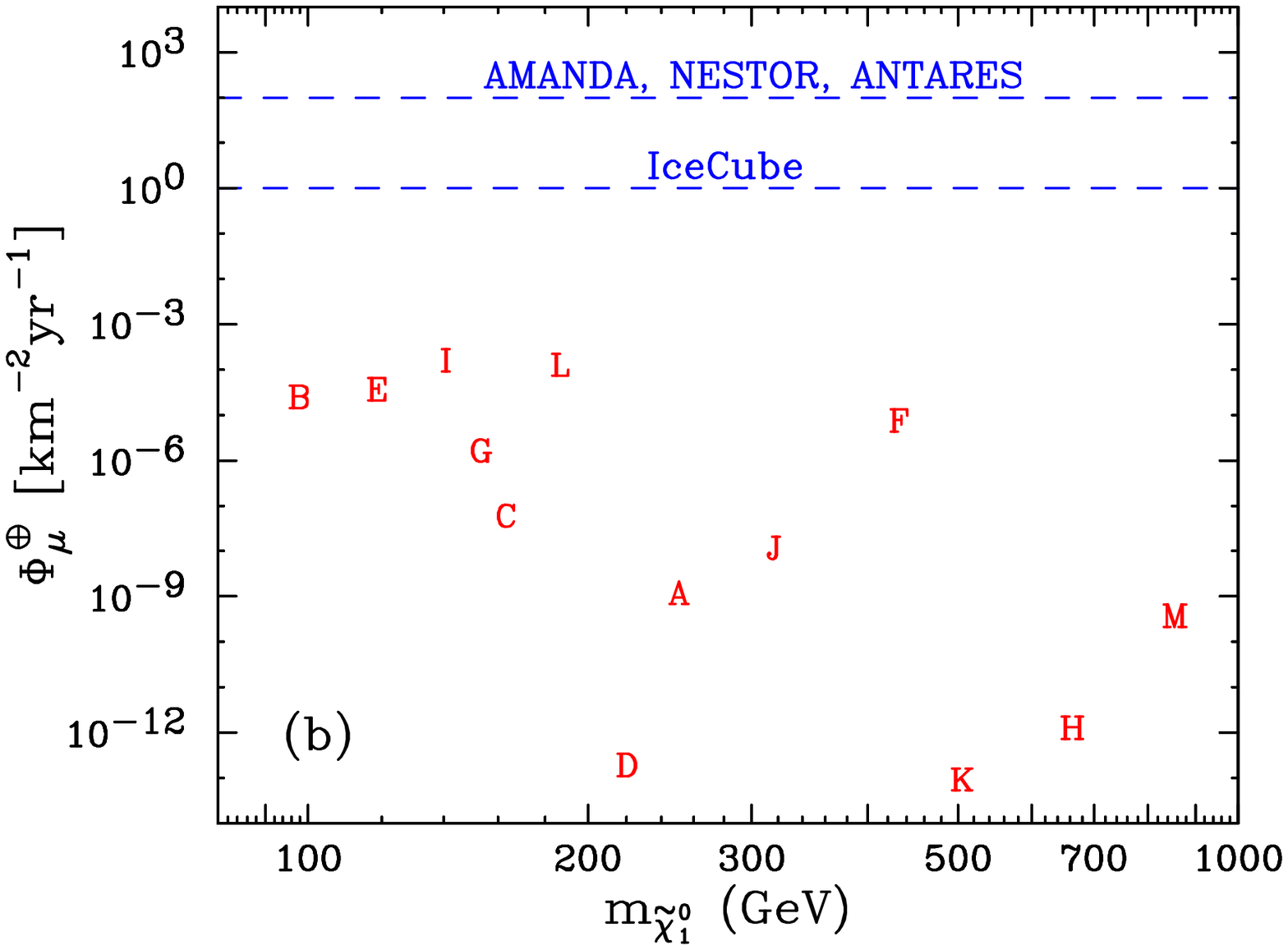}{0.99}
\end{minipage}
\caption{\label{fig:muons} 
{\it Muon fluxes from neutrinos originating from relic annihilations
inside (a) the Sun and (b) the Earth. Approximate sensitivities of
near future neutrino telescopes ($\Phi_{\mu} = 10^2~\km^{-2}~\yr^{-1}$
for AMANDA II~\cite{AMANDA}, NESTOR~\cite{NESTOR}, and
ANTARES~\cite{ANTARES}, and $\Phi_{\mu} = 1~\km^{-2}~\yr^{-1}$ for
IceCube~\cite{IceCube}) are also indicated.  }}
\end{figure}

Muon fluxes for each of the benchmark points are given in
Fig.~\ref{fig:muons}, using {\tt Neutdriver} with a fixed constant
local density $\rho_0 = 0.3~\gev/\cm^3$ and neutralino velocity
dispersion $\bar{v} = 270~\km/\s$.  For the points considered, rates
from the Sun are far more promising than rates from the Earth.  For
the Sun, muon fluxes are for the most part anti-correlated with
neutralino mass for the reason noted above.  There are two strong
exceptions, however: the focus point models E and F have anomalously
large fluxes.  In these cases, the dark matter's Higgsino content,
though still small, is significant (see Table~\ref{table:I}), leading
to annihilations to gauge boson pairs, hard neutrinos, and enhanced
detection rates, as discussed above.

The potentials of current and planned neutrino telescopes have been
reviewed in~\cite{Feng:2001zu}. The exact reach depends on the salient
features of a particular detector, {\em e.g.}, its physical dimensions
and muon energy threshold, and the expected characteristics of the
signal, {\em e.g.}, its angular dispersion, energy spectrum and source
(Sun or Earth).  Two sensitivities, which are roughly indicative of
the potential of upcoming neutrino telescope experiments, are given in
Fig.~\ref{fig:muons}. For focus-point model E, where the neutralino is
both light and significantly different from pure Bino-like, detection
in the near future at AMANDA II~\cite{AMANDA}, NESTOR~\cite{NESTOR},
and ANTARES~\cite{ANTARES} is possible.  Point F may be within reach
of IceCube~\cite{IceCube}, as the neutralino's significant Higgsino
component compensates for its large mass.  For point B, and possibly
also points I, G, C, and L, the neutralino is nearly pure Bino, but is
sufficiently light that detection at IceCube may also be possible.

Muon energy thresholds specific to individual detectors have not been
included.  For AMANDA II and, especially, IceCube, these thresholds
may be large, significantly suppressing the muon signal in models with
$\mchi$ less than about 4 to 6 $E_{\mu}^{\rm
th}$~\cite{Bergstrom:1997tp,Barger:2001ur}.  Note also that, for
certain neutralino masses and properties, a population of dark matter
particles in solar system orbits may boost the rates presented here by
up to two orders of magnitude~\cite{Damour:1998rh}.  While this effect
deserves further study, here we have conservatively neglected this
possible enhancement.

\section{Photons from Annihilations in the Galactic Center}
\label{sec:photons}

As with the centers of the Sun and Earth, the center of the galaxy may
attract a significant overabundance of relic dark matter
particles~\cite{Urban:1992ej}.  Relic pair annihilation at the
galactic center will then produce an excess of photons, which may be
observed in gamma ray detectors.  While monoenergetic signals from
$\chi \chi \rightarrow \gamma \gamma$ and $\chi \chi \rightarrow
\gamma Z$ would be spectacular~\cite{Bergstrom:1998fj}, they are
loop-suppressed and unobservable for these benchmark points. We
therefore consider continuum photon signals here.

The integrated photon flux above some photon energy threshold $\ethr$
is~\cite{Feng:2001zu}
\begin{equation}
\Phi_{\gamma} (\ethr)=  5.6 \times 10^{-10}~\cm^{-2}~\s^{-1}
\times \sum_i \int_{\ethr}^{\mchi} \! \! dE \frac{dN_{\gamma}^i}{dE}
\left( \frac{\sigma_i v}{{\rm pb}} \right)
\left( \frac{100~\gev}{\mchi} \right)^2
\bar{J}(\Delta \Omega) \, \Delta \Omega \ ,
\label{phigamma}
\end{equation}
where the sum is over all annihilation channels $i$,
$dN_{\gamma}^i/dE$ is the differential gamma ray multiplicity for
process $i$, $\Delta \Omega$ is the solid angle of the field of view
of a given telescope, and $\bar{J}$ is a measure of the cuspiness of
the galactic halo density profile.  There is a great deal of
uncertainty in $\bar{J}$, with possible values in the range 3 to
$10^5$~\cite{Bergstrom:1998fj}.

\begin{figure}[tb]
\postscript{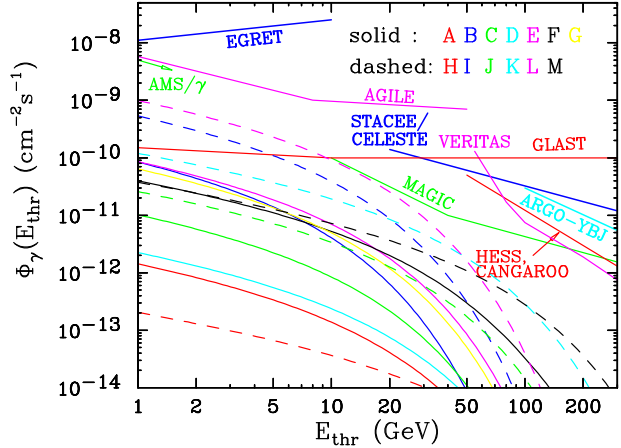}{0.49}
\caption{\label{fig:photon_spectra} 
{\it The integrated photon flux $\Phi(\ethr)$ as a function of photon
energy threshold $\ethr$ for photons produced by relic annihilations
in the galactic center. A moderate halo parameter $\bar{J} = 500$ is
assumed.  Point source flux sensitivities for various gamma ray
detectors are also shown. }}
\end{figure}

The integrated photon flux $\Phi(\ethr)$ is given in
Fig.~\ref{fig:photon_spectra} for each of the benchmark points.  We
choose $\Delta \Omega = 10^{-3}$ and a moderate value of $\bar{J} =
500$.  Estimates for point source flux sensitivities of several gamma
ray detectors, both current and planned, are also shown.  The
space-based detectors EGRET, AMS/$\gamma$ and GLAST can detect soft
photons, but are limited in flux sensitivity by their small effective
areas.  Ground-based telescopes, such as MAGIC, HESS, CANGAROO and
VERITAS, are much larger and so sensitive to lower fluxes, but are
limited by higher energy thresholds.  These sensitivities are not
strictly valid for observations of the galactic center.  Nevertheless,
they provide rough guidelines for what sensitivities may be expected
in coming years.  For a discussion of these estimates, their
derivation, and references to the original literature,
see~\cite{Feng:2001zu}.

\begin{figure}[tb]
\begin{minipage}[t]{0.49\textwidth}
\postscript{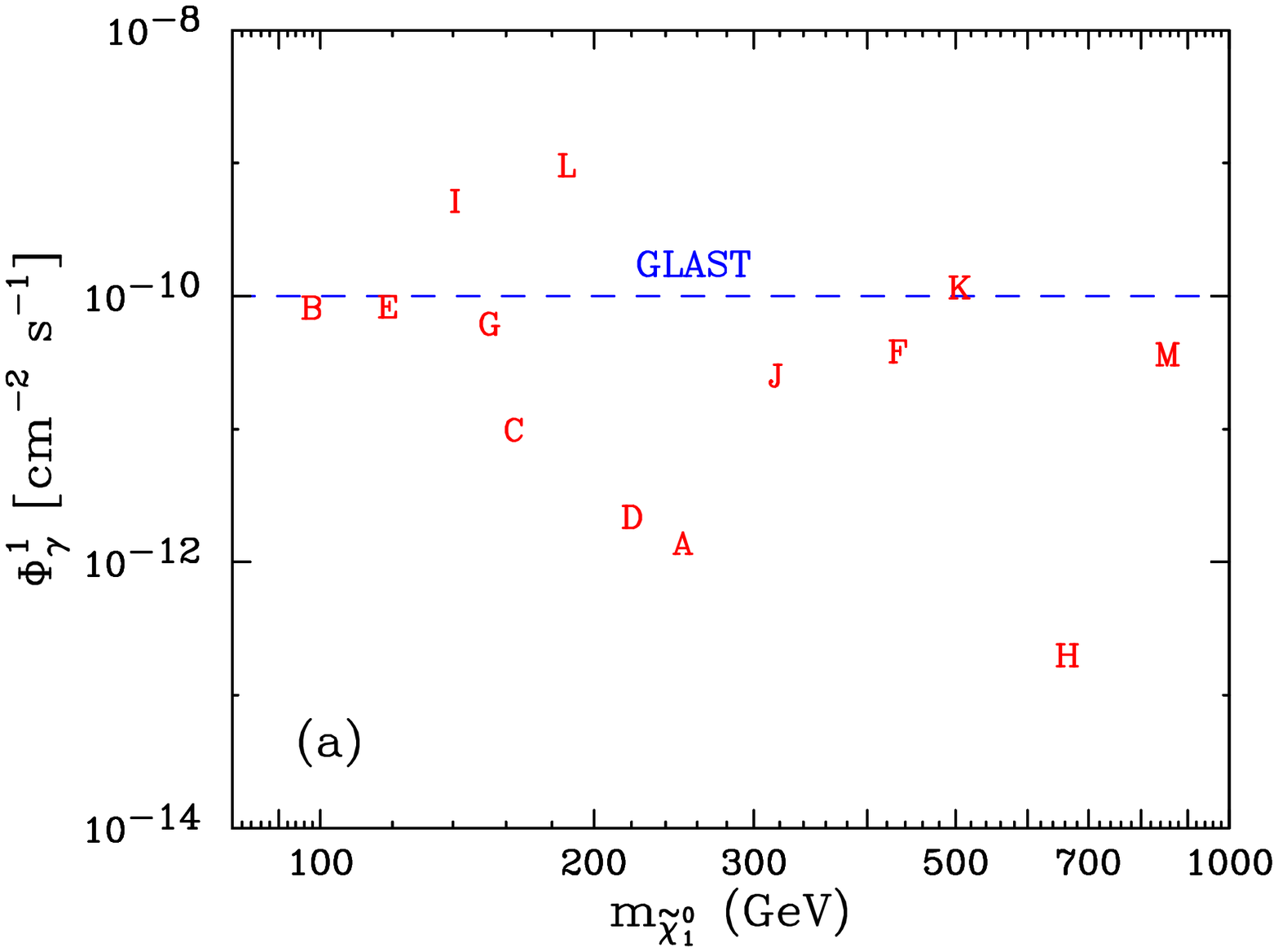}{0.99}
\end{minipage}
\hfill
\begin{minipage}[t]{0.49\textwidth}
\postscript{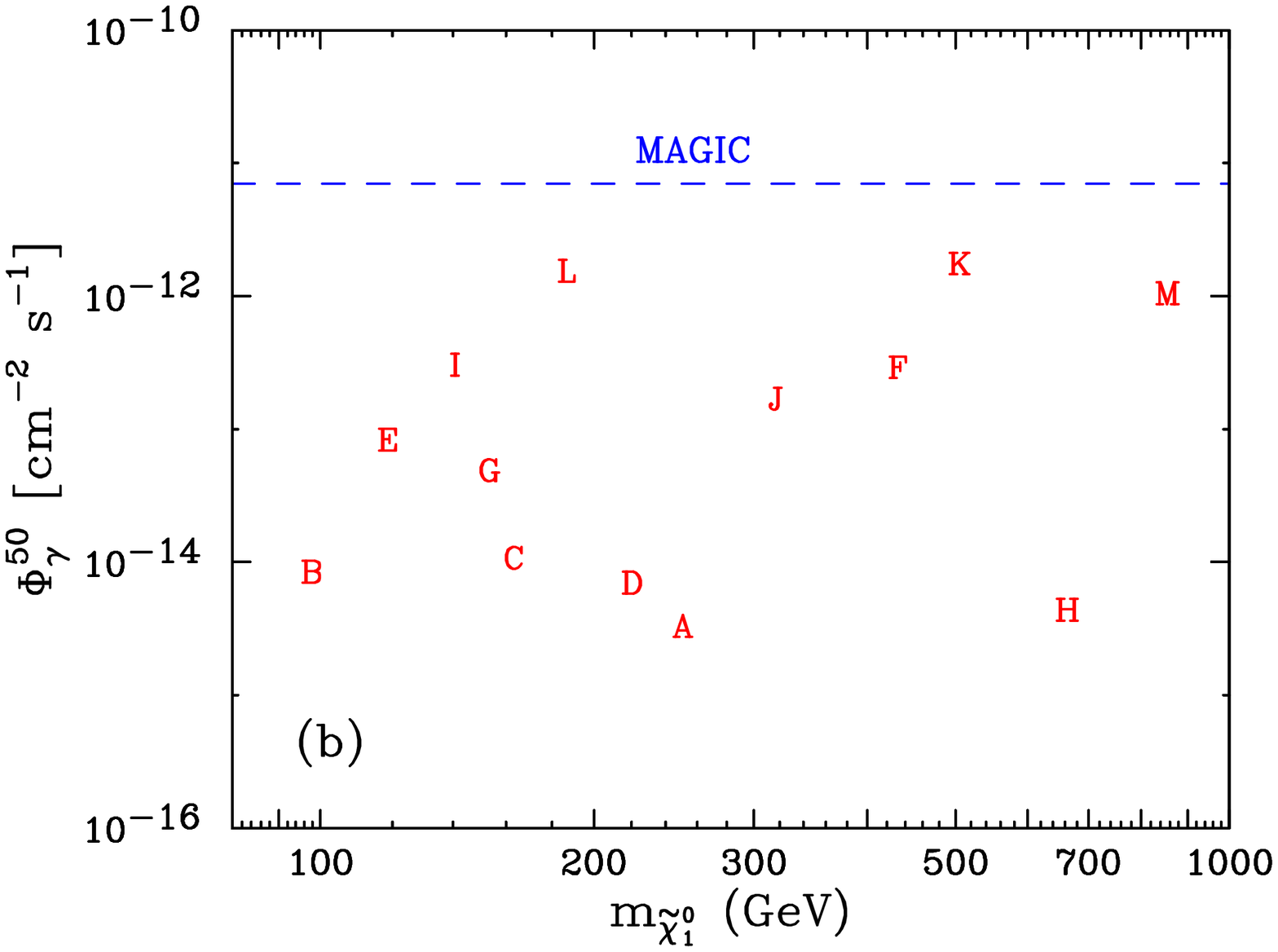}{0.99}
\end{minipage}
\caption{\label{fig:photons} 
{\it Comparisons between predicted integrated fluxes and prospective
experimental sensitivities for photons with (a) a 1~GeV threshold, and
(b) a 50~GeV threshold, following~\cite{Feng:2001zu}. Estimated
sensitivities for (a) GLAST~\cite{GLAST} and (b) MAGIC~\cite{MAGIC}
are also shown.  A moderate halo parameter $\bar{J}=500$ is
assumed. }}
\end{figure}

Integrated fluxes for the benchmark points are given in
Fig.~\ref{fig:photons} for two representative energy thresholds:
1~GeV, accessible to space-based detectors, and 50~GeV, characteristic
of ground-based telescopes.  Estimated sensitivities for two of the
more promising experiments, GLAST~\cite{GLAST} and MAGIC~\cite{MAGIC},
are also shown.  {}From (\ref{phigamma}), we expect the photon flux to
be inversely correlated with neutralino mass.  Roughly speaking, this
general trend is seen in Fig.~\ref{fig:photons}a. For
Fig.~\ref{fig:photons}b, it is offset by the requirement of a hard
photon, which suppresses the signal from light neutralinos.  In both
cases, however, this general trend may be disrupted by a variety of
additional effects.  In particular, the photon spectrum is relatively
hard for annihilation to gauge bosons; $\Phi(\ethr)$ is, then,
enhanced for the focus point models E and F, which have neutralinos
with significant Higgsino components.  The cross section $\sigma_i$
for annihilation to $b\bar{b}$ through $s$-channel pseudoscalar Higgs
is also enhanced for large $\tan\beta$~\cite{Feng:2000gh}, boosting
photon signals at points I, J, K, L, and M.

GLAST appears to be particularly promising, with points I and L giving
observable signals.  Recall, however, that all predicted fluxes scale
linearly with $\bar{J}$.  For isothermal halo density profiles, the
fluxes may be reduced by two orders of magnitude.  On the other hand,
for particularly cuspy halo models, such as those
in~\cite{Navarro:1996iw}, all fluxes may be enhanced by two orders of
magnitude, leading to detectable signals in GLAST for almost all
points, and at MAGIC for the majority of benchmark points.

\section{Positrons from Annihilations in the Galactic Halo}
\label{sec:positrons}

\begin{figure}[tb]
\postscript{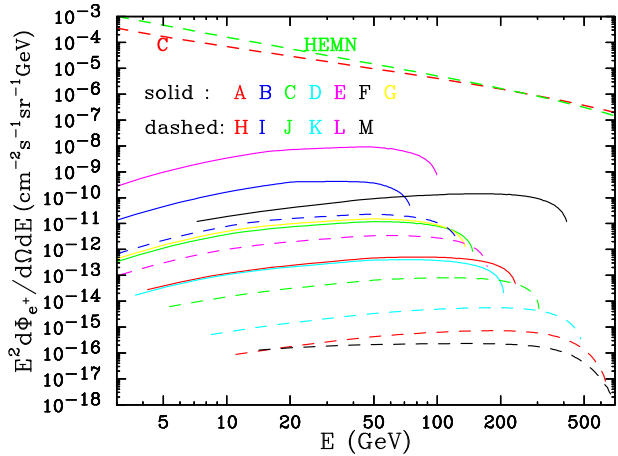}{0.49}
\caption{\label{fig:positron_spectra} 
{\it Differential positron fluxes produced by relic annihilations in
the galactic halo.  Background fluxes are also shown for two models
from~\cite{Moskalenko:1999sb}.  }}
\end{figure}

Relic neutralino annihilations in the galactic halo may also be
detected through positron excesses in space-based and balloon
experiments~\cite{Tylka:1989xj,Moskalenko:1999sb}.  The positron flux
may be written as~\cite{Moskalenko:1999sb}
\begin{equation}
\frac{d\Phi_{e^+}}{d\Omega dE} = \frac{\rho^2}{\mchi^2}
\sum_i \sigma_i v B_{e^+}^i
\int dE_0\, f_i(E_0)\, G(E_0, E) \ ,
\label{dPhi}
\end{equation}
where $\rho$ is the local neutralino mass density, the sum is over all
annihilation channels $i$, and $B_{e^+}^i$ is the branching fraction
to positrons in channel $i$. The initial positron energy distribution
is given by the source function $f(E_0)$, and the Green function
$G(E_0, E)$ propagates positrons in the galaxy.  We use the Green
function corresponding to a modified isothermal halo with size 4 kpc
given in~\cite{Moskalenko:1999sb}.  The differential positron fluxes
for the benchmark points are given in Fig.~\ref{fig:positron_spectra}.
Note that the background spectrum drops rapidly with energy; hard
positrons from neutralino annihilation are most easily observed.

To estimate the observability of a positron excess, we follow the
procedure advocated in~\cite{Feng:2001zu}. For each benchmark
spectrum, we find the positron energy $\eopt$ at which the positron
signal to background ratio $S/B$ is maximized.  For detection, we then
require that $S/B$ at $\eopt$ be above some value.  The sensitivities
of a variety of experiments have been estimated in~\cite{Feng:2001zu}.
Among these experiments, the most promising is AMS~\cite{AMS}, the
anti-matter detector to be placed on the International Space Station.
AMS will detect unprecedented numbers of positrons in a wide energy
range.  We estimate that a 1\% excess in an fairly narrow energy bin,
as is characteristic of the neutralino signal, will be statistically
significant.

\begin{figure}[tb]
\begin{minipage}[t]{0.49\textwidth}
\postscript{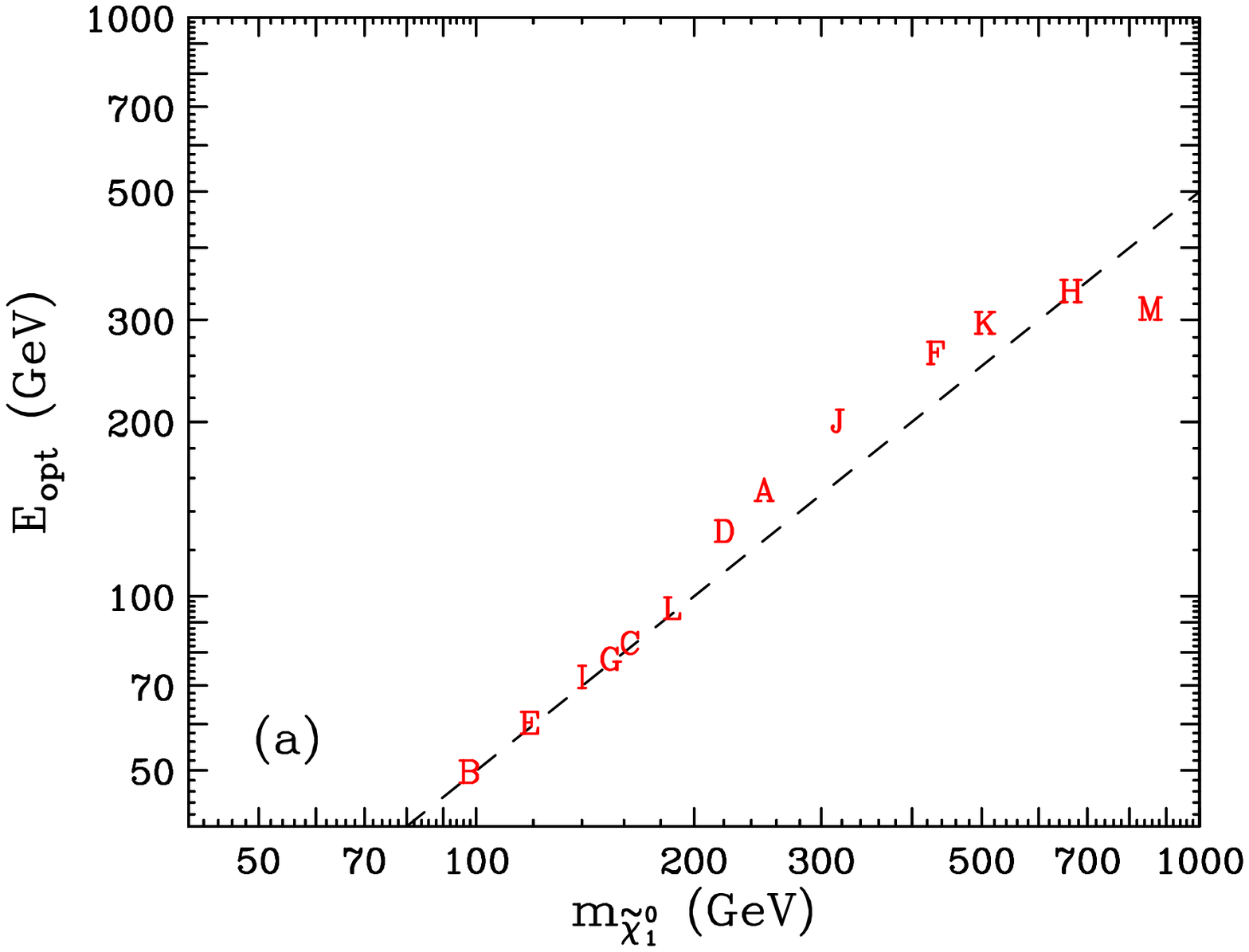}{0.99}
\end{minipage}
\hfill
\begin{minipage}[t]{0.49\textwidth}
\postscript{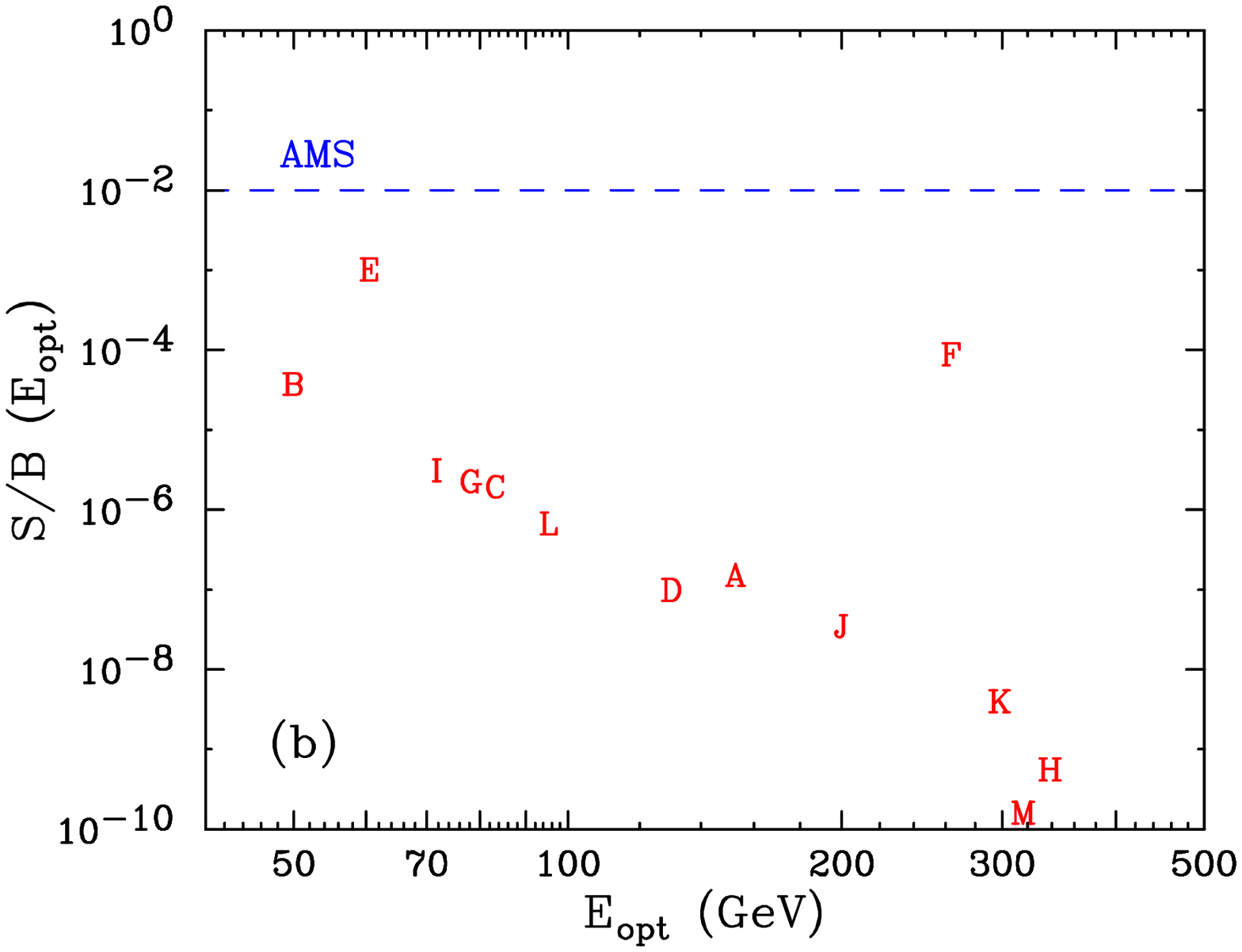}{0.99}
\end{minipage}
\caption{\label{fig:positrons} 
{\it (a) The optimal energies for which the positron signal to
background ratios $S/B$ are maximized, and (b) $S/B$ at these energies
for each of the benchmark points, following~\cite{Feng:2001zu}.  In
(a), the dashed line is for $\eopt = \mchi/2$, and in (b), the
estimated sensitivity of the AMS~\cite{AMS} experiment is shown.  }}
\end{figure}

Estimates of $\eopt$ and the maximal $S/B$ for each benchmark point
are given in Fig.~\ref{fig:positrons}.  To an excellent approximation,
energetic positrons are produced only when neutralinos annihilate to
gauge bosons that decay directly to positrons.  Because this decay is
two-body, $\eopt \approx \mchi/2$ for all benchmark points.  As
expected from (\ref{dPhi}), $S/B$ is typically inversely correlated
with neutralino mass.  As discussed in Sec.~\ref{sec:neutrinos} for
the case of neutrinos, however, there are two strong exceptions: the
focus point models E and F.  Again, these points have mixed
gaugino-Higgsino dark matter. Rates for annihilation to gauge bosons
and, consequently, the positron signals are therefore greatly
enhanced.

Even for points E and F, however, discovery of the positron excess is
challenging for the smooth isothermal halo considered here.  The
positron search is most effective for light neutralinos that are more
Higgsino-like than those represented in this set of benchmark
points. However, as with the photon signal, positron rates are
sensitive to the halo model assumed; for clumpy
halos~\cite{Silk:1992bh}, the rate may be enhanced by orders of
magnitude~\cite{Moskalenko:1999sb}.

\section{Conclusions}
\label{sec:conclusions}

In this paper, we have provided indicative estimates of the rates that
could be expected for the benchmark supersymmetric scenarios proposed
in~\cite{Battaglia:2001zp}. We emphasize that, in addition to the
supersymmetric model dependences of these calculations, there are
important astrophysical uncertainties. These include the overall halo
density, the possibility that it may be enhanced in the solar system,
its cuspiness near the galactic centre, and its clumpiness
elsewhere. For these reasons, our conclusions about the relative ease
with which different models may be detected using the same signature
may be more reliable than the absolute strengths we predict, or
comparisons between the observabilities of different
signatures. Nevertheless, our estimates do indicate that there may be
good prospects for astrophysical detection of quite a large number of
the benchmark scenarios.

In particular, the direct detection of relic particles by
spin-independent elastic scattering in models I, B, E and L may be
possible using the projected GENIUS~\cite{GENIUS} detector, with
models G, F and C not far from the likely threshold of
detectability. The prospects of detecting spin-dependent elastic
scattering do not, however, look so promising in the benchmark
scenarios studied. The indirect detection of muons generated by
high-energy neutrinos due to annihilations inside the Sun should be
most easily detectable in models E, F and B, followed by models I, G,
L and C, which offer prospects with the proposed
IceCube~\cite{IceCube} detector.  However, unless there is a
substantial solar-system enhancement, the prospects for detecting
annihilations inside the Earth are not so encouraging.  Models L and I
offer the best prospects for the detection of photons from
annihilations in the galactic centre, followed by models K, B, E and
G.  Here the best prospects may be those for the GLAST~\cite{GLAST}
satellite, with its relatively low threshold. However, there may also
be prospects for ground-based experiments such as MAGIC~\cite{MAGIC},
if the halo is cuspier at the galactic centre than we have
assumed. Models E, F and B offer the positron signals with the largest
signal-to-background ratios, though apparently requiring a sensitivity
greater than that expected for AMS~\cite{AMS}, unless the halo is
rather clumpy.

In specifying the benchmark models, the constraint coming from the
anomalous magnetic moment of the muon was not imposed
rigorously. However, it was noted that the more $g_\mu - 2$-friendly
models I, L, B, G, C and J offered good prospects for detecting
several supersymmetric particles at the LHC and/or a linear $e^+ e^-$
collider with 1~TeV in the centre of mass. Most of these models also
exhibit good prospects for dark matter detection, with the exception
of model J. Among the less $g_\mu - 2$-friendly models, we note that
E, F and K offer some astrophysical prospects. This is particularly
interesting in the case of focus point model F, which does not offer
generous prospects at colliders, and model K, which is not easy to
explore with a linear $e^+ e^-$ collider. On the other hand, models M
and H, which are particularly difficult to explore with colliders,
also do not offer bright prospects for astrophysical detection.

Our analysis indicates the effort required to cover the possible
supersymmetric parameter space via a number of different astrophysical
signatures, at least within the CMSSM assumptions used here. It would
be interesting to extend such a benchmark analysis to other types of
supersymmetric models, but that lies beyond the scope of this
paper. Ultimately, one would hope to be able to confront accelerator
and astrophysical measurements of supersymmetry, and make non-trivial
cross-checks of our CMSSM assumptions, but that is for the future. For
the moment, the race to discover supersymmetry is still open, and our
analysis indicates that there may be good prospects for detecting
supersymmetric dark matter before the LHC comes into operation.

\vskip 0.5in
\vbox{
\noindent{ {\bf Acknowledgments} } \\
\noindent  
The work of J.L.F. was supported in part by the US Department of
Energy under cooperative research agreement DF--FC02--94ER40818.  The
work of K.A.O. was supported partly by DOE grant
DE--FG02--94ER--40823.  K.T.M. thanks the Fermilab Theory Group for
hospitality during the completion of this work.}

\end{document}